# LOTKA'S INVERSE SQUARE LAW OF SCIENTIFIC PRODUCTIVITY: ITS METHODS AND STATISTICS

Stephen J. Bensman

LSU Libraries (Retired)

Louisiana State University

Baton Rouge, LA 70803 USA

E-mail: notsjb@lsu.edu

Lawrence J. Smolinsky

Department of Mathematics

Louisiana State University

Baton Rouge, LA 70803 USA

E-mail: smolinsk@math.lsu.edu

**Abstract**

This brief communication analyzes the statistics and methods Lotka used to derive his inverse square law of scientific productivity from the standpoint of modern theory. It finds that he violated the norms of this theory by extremely truncating his data on the right. It also proves that Lotka himself played an important role in establishing the commonly used method of identifying power-law behavior by the R^2 fit to a regression line on a log-log plot that modern theory considers unreliable by basing the derivation of his law on this very method.

## Introduction

In recent years power-law distributions as scientific models have come under intensive scrutiny. Of primary importance in this have been the papers by Newman (2005) and Clauset, Shalizi, and Newman (2009). In his paper Newman (2005, p. 323) states that when the probability of measuring a particular value of some quantity varies inversely as a power of that value, the quantity is said to follow a power law. Lotka's Inverse Square Law of Scientific Productivity—historically the first law of scientometrics—is precisely such a law. Power-law distributions have the general shape of negative exponential J-curves with a long tail to the right, and a key characteristic of them is a surfeit of observations at the right tip of what is termed this "heavy tail." Clauset, Shalizi, and Newman (2009) focus on the problems of identifying power-law distributions, and they aver that "the detection and characterization of power laws is complicated by the large fluctuations that occur in the tail of the distribution—the part of the distribution representing large but rare events—and by the difficulty of identifying the range over which power-law behavior holds" (p. 661). They also assert that the commonly used method of identifying power-law distributions by logging the variables on both axes of the graph and then using regression analysis to measure the R^2 linear fit to the resulting regression line or trendline is unreliable. It will be seen in the subsequent analysis of Lotka's methods and data that his inverse square law of scientific productivity suffers from all these problems and that Lotka himself used the method of the linear fit on the log-log plot to derive his law of scientific productivity.

## Lotka's Law

Lotka's Inverse Square Law of Scientific Productivity is eponymously named after Alfred J. Lotka, and it is the first scientometric or informetric law. To obtain the data for deriving his law, Lotka (1926) made a count of the number of personal names in the 1907-1916 decennial index of *Chemical Abstracts* against which there appeared 1, 2, 3, etc. entries, covering only the letters A and B of the alphabet. He also applied a similar process to the name index in Felix Auerbach's *Geschichtstafeln der Physik (*Leipzig: J. A. Barth, 1910), which dealt with the entire range of the history of physics through 1900. By using the latter source, Lotka hoped to take into account not only the volume of production but also quality, since it listed only the outstanding contributions in physics. In making these counts Lotka credited only the senior author in joint publications. On the basis of this data, Lotka derived what he termed an “inverse square law“, according to which of any set of authors, ca. 60% produce one paper, whereas the percent producing 2 is 1 /2^2 or ca. 25%, the percent producing 3 equals 1/ 3^2 or ca. 11.1%, the percent producing 4 is 1/ 4^2 or ca. 6.3%, etc. Thus, of 1000 authors, 600 produce 1 paper, 250 produce 2 papers, 111 produce 3 papers, and 63 produce 4 papers. Lotka (1926, p. 320) defined the general formula for the relation he found between the frequency of *y* of persons making *x* contributions as: $x^n y$ = const. Most interestingly, Lotka (1926, p. 320) referred to the exponent as the “slope,” and he stated that, as determined by “least squares,” the “slope of the curve” to his *Chemical Abstracts* data was found to be 1.888 and to his Auerbach data 2.021. The source of these statements can be located in Figure 1 below, which replicates the two histograms found in his article. Part A presents the percent chemists and physicists on the y-axis and the number of mentions up to 10 on the x-axis. However, on the histogram in Part B both the axes are logged, and the histogram is designed to show closeness of the linear fit to resulting regression lines determined by least-squares analysis. It will be seen below that the absolute values of the slopes of these regression lines are equal to the exponents. Thus, Lotka played an important role in establishing the method of identifying power-law behavior by the linear fit to a regression line on a log-log plot found unreliable by Clauset, Shalizi, and Newman (2009) by basing his derivation of his law on this method. It will also be shown that Lotka avoided one of the main problems identified by them for fitting data to power laws—the extreme outliers on the fat tail—by truncating his data on the right.

**Figure 1. Lotka's graphs of his inverse square power law of scientific productivity**

A. Histogram showing % of authors mentioned once, twice, etc.

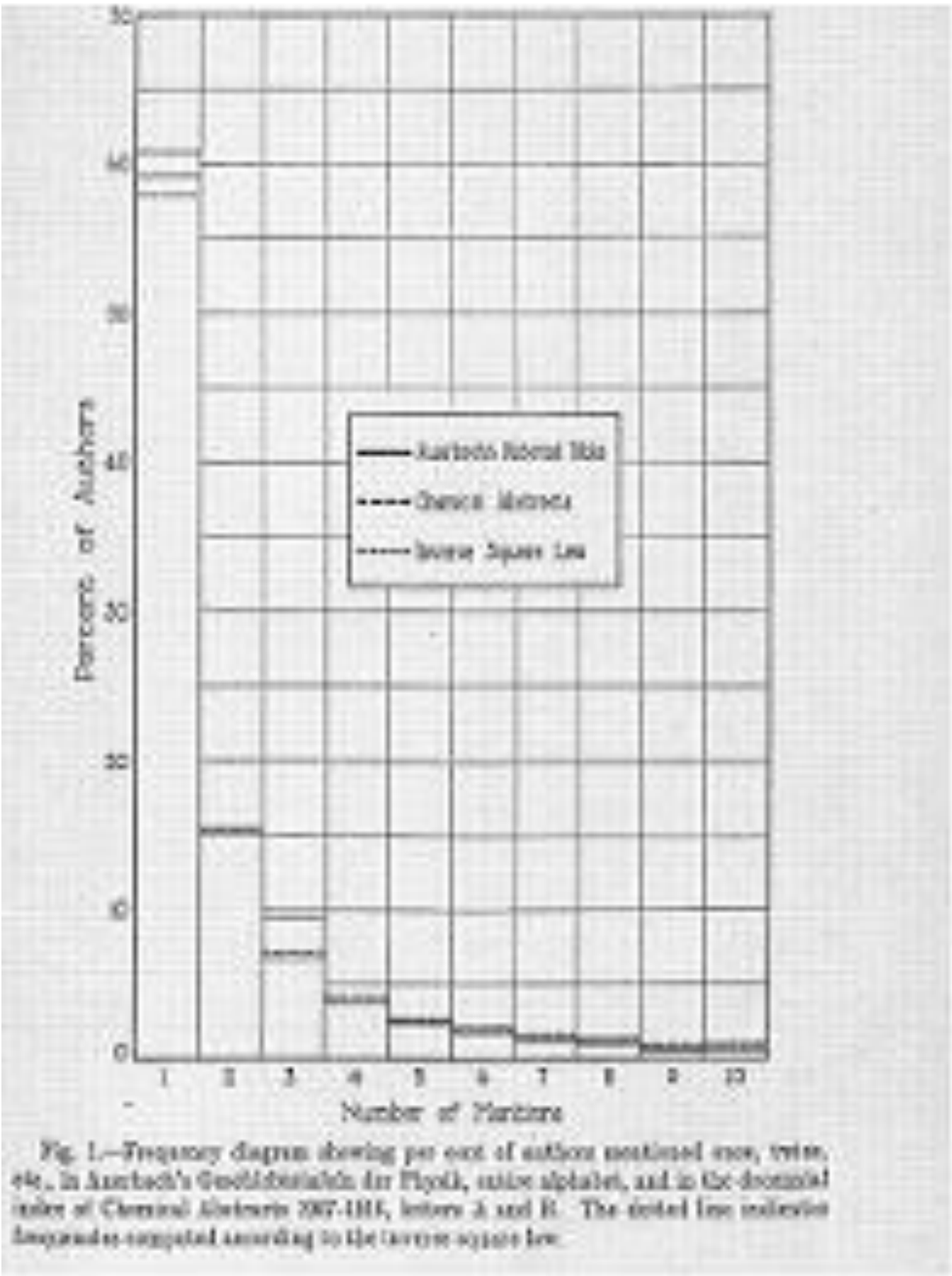

B. Logarithmic histogram showing number of authors mentioned once, twice, etc.

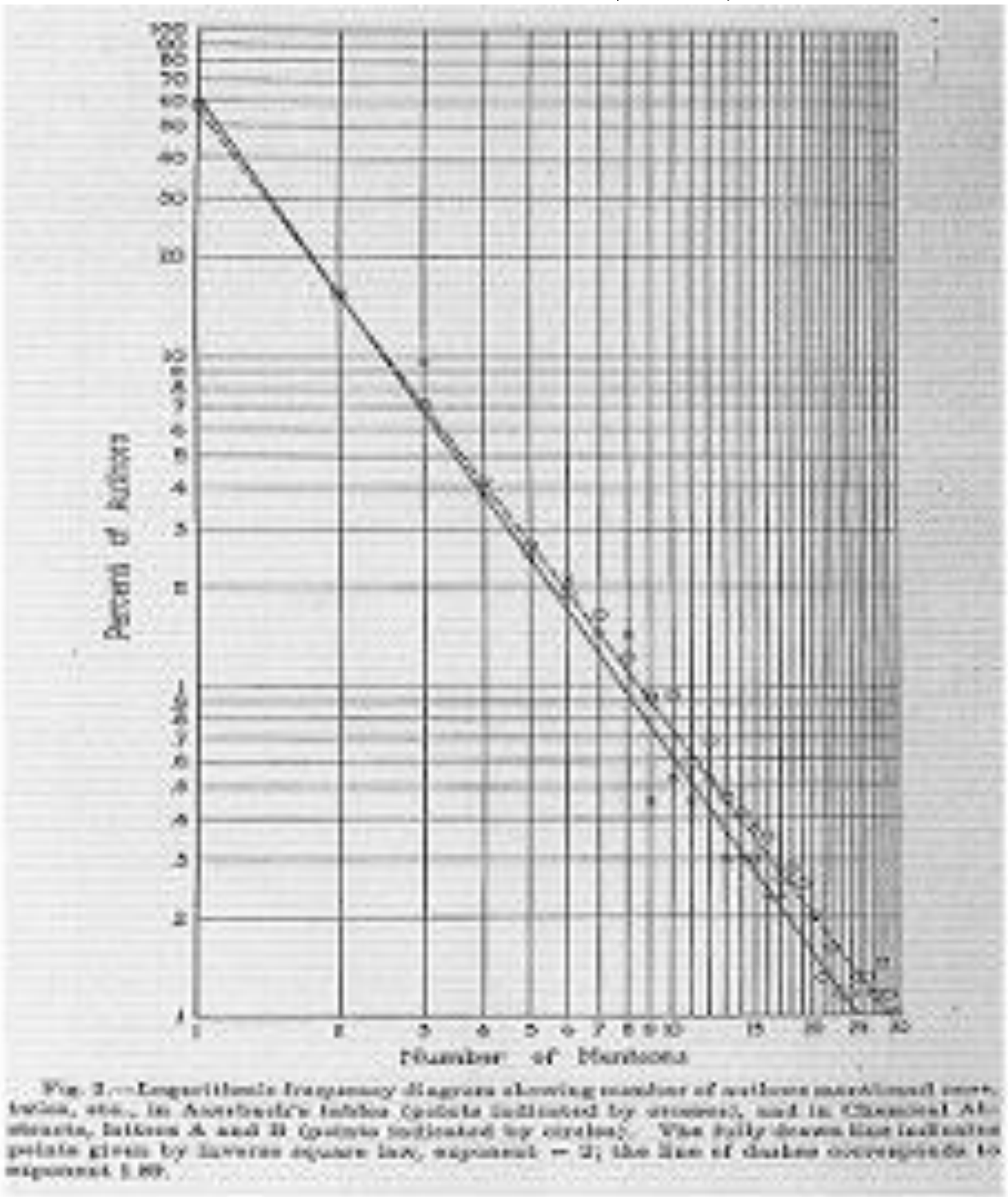

Note: From "The frequency distribution of scientific productivity," by A. J. Lotka, 1926. *Journal of the Washington Academy of Sciences*, 16 (12), pp. 321-322.

**Lotka's Right Truncation**

At the beginning of his paper Newman (2005, p. 325) took up the problem of extreme outliers on the right of the distribution in the linear fit to the regression line on the log-log plot. He noted that often the fit is not a good one because the right-hand end of the distribution, where the outliers are located, is noisy due to sampling errors as a result of the wide spacing of the observations there. He observed that one way to deal with this problem would be to throw out the data on the tail of the curve. But he advised against this, because "…there is often useful information in those data and …many distributions follow a power law only in the tail, so we are in danger of throwing out the baby with the bathwater" (p. 325). However, as now will be shown, this is precisely what Lotka did.

The range of works for Lotka's *Chemical Abstracts* data ran from 1 to 346 and for his Auerbach data from 1 to 48. However, Lotka decided to truncate the range for the *Chemical Abstracts* distribution at 30 works and for the Auerbach data at 17. The reason for this Lotka (1926) gave in a footnote, where he stated that beyond these points "fluctuations become excessive owing to the limited number of persons in the sample" (p. 320n). In another footnote (p. 323n) Lotka stated that fortunately there were somewhat more persons of very great productivity than would be expected under his simple law, calling attention to the extreme outlier at 346 works in his *Chemical Abstracts* data—the Swiss biochemist, Emil Abderhalden. Lotka noted that such outliers should perhaps be considered separately because they are not the product of one person due to his policy of counting only senior authors of joint works. Although expressed in a footnote these considerations are extremely portentous for the following reasons: 1) Lotka appeared to be aware of the distortions that could result from right truncation that underlie Newman's admonition against this procedure; 2) he seemed to have intuited that a key characteristic of the power-law model he was pioneering was an excess of "large but rare events" over what could be generally expected; and 3) in measuring productivity a complexity is that a single "item" (e.g., an article) could have multiple "sources" (e.g., scientists), making the proper allocation of credit for a given article to a given scientist a hellish process.

| **Table 1. Full distributions of Lotka's *Chemical Abstracts* and Auerbach data** | | |
|---|---|---|
| ***Number f Authors*** | ***Range of Works*** | ***Number of Works*** |
| ***Chemical Abstracts* Data** | | |
| 6891 | 1-346 | 22934 |
| **Auerbach Data** | | |
| 1325 | 1-48 | 3398 |

| **Table 2. Scale of Lotka's right truncation** | | | | | |
|---|---|---|---|---|---|
| ***Number of Authors*** | ***Percent of Authors*** | ***Range of Works*** | ***Percent Range of Works*** | ***Number of Works*** | ***Percent Number of Works*** |
| ***Chemical Abstracts* Data** | | | | | |
| 0 | 0.00% | 316 | 91.33% | 3818 | 16.65% |
| **Auerbach Data** | | | | | |
| 0 | 0.00% | 31 | 64.58% | 451 | 13.27% |

Tables 1 and 2 above set forth the results of Lotka's right truncation. In both cases he did not cut the number of authors but did his calculations in terms of the total number of authors in the full distributions. Through this method Lotka partially incorporated the probability structure of the full distributions into his calculations based on the truncated distributions. As for range and number of works, Lotka reduced these respectively by 91.33% and 16.65% in his *Chemical Abstracts* data and respectively by 64.58% and 13.27% for his Auerbach data. The histograms in Figures 2 and 3 below graph the amount of truncation both in term of numbers (Part A) and percentages (Part B). Here the bins were set to equal or approximate one-half the amount of truncation—15 for *Chemical Abstracts* and 8 for Auerbach. Therefore everything to the right of the second bin was truncated except the number of authors, which remained the same for both the full and truncated distributions.

**Lotka's Method of Calculating the R^2 Fit to the Regression Line on the Log-Log Plot**

While Lotka did not calculate R^2, he used ordinary least squares to fit his data. Ordinary least squares calculates the trendline by minimizing the residual sum of squares, but, for fixed data, it is.

**Figure 2. Full frequency distributions of Lotka's *Chemical Abstracts* data: each bin being defined by being one-half of Lotka's right-truncation point of 30**

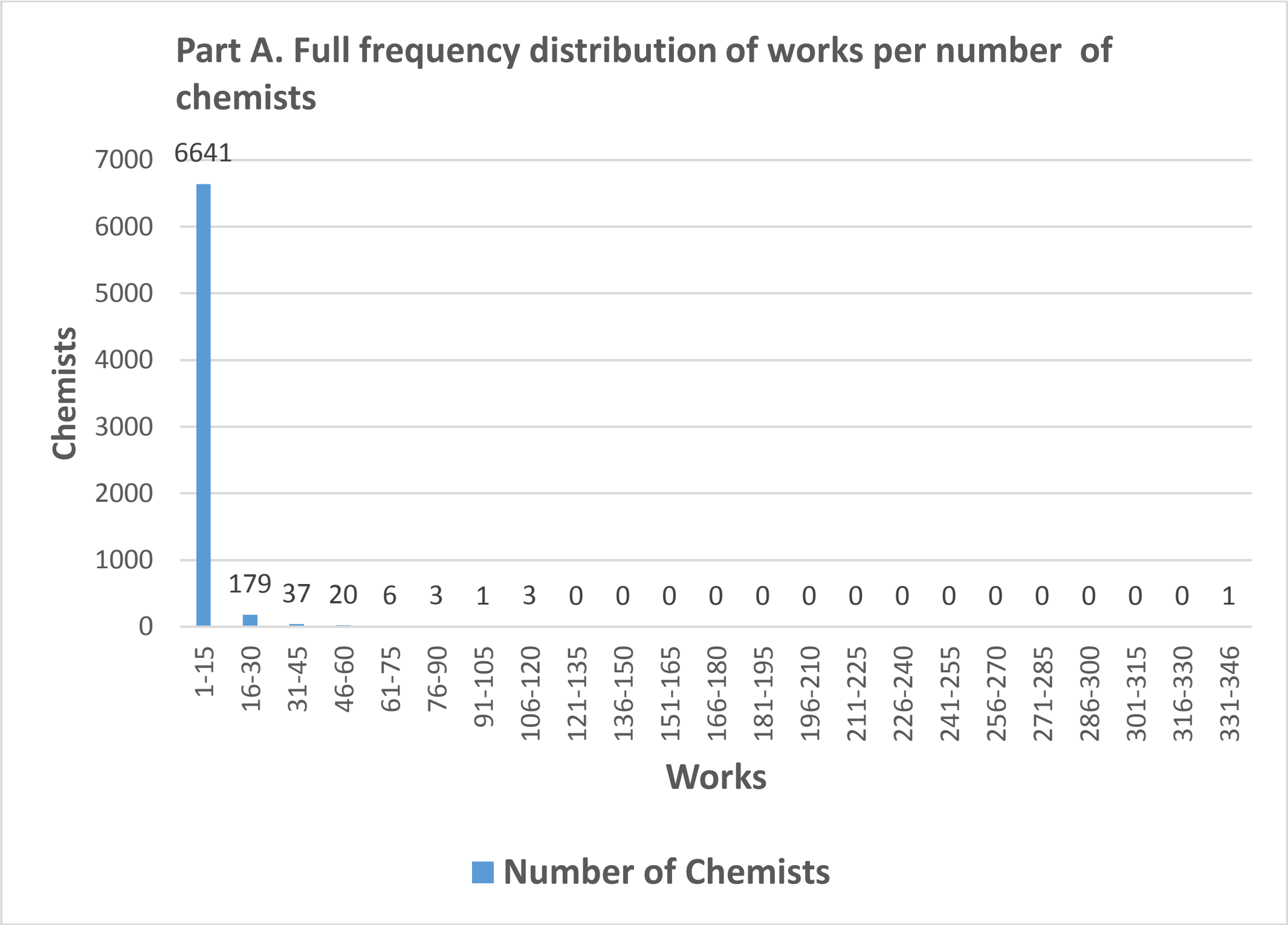

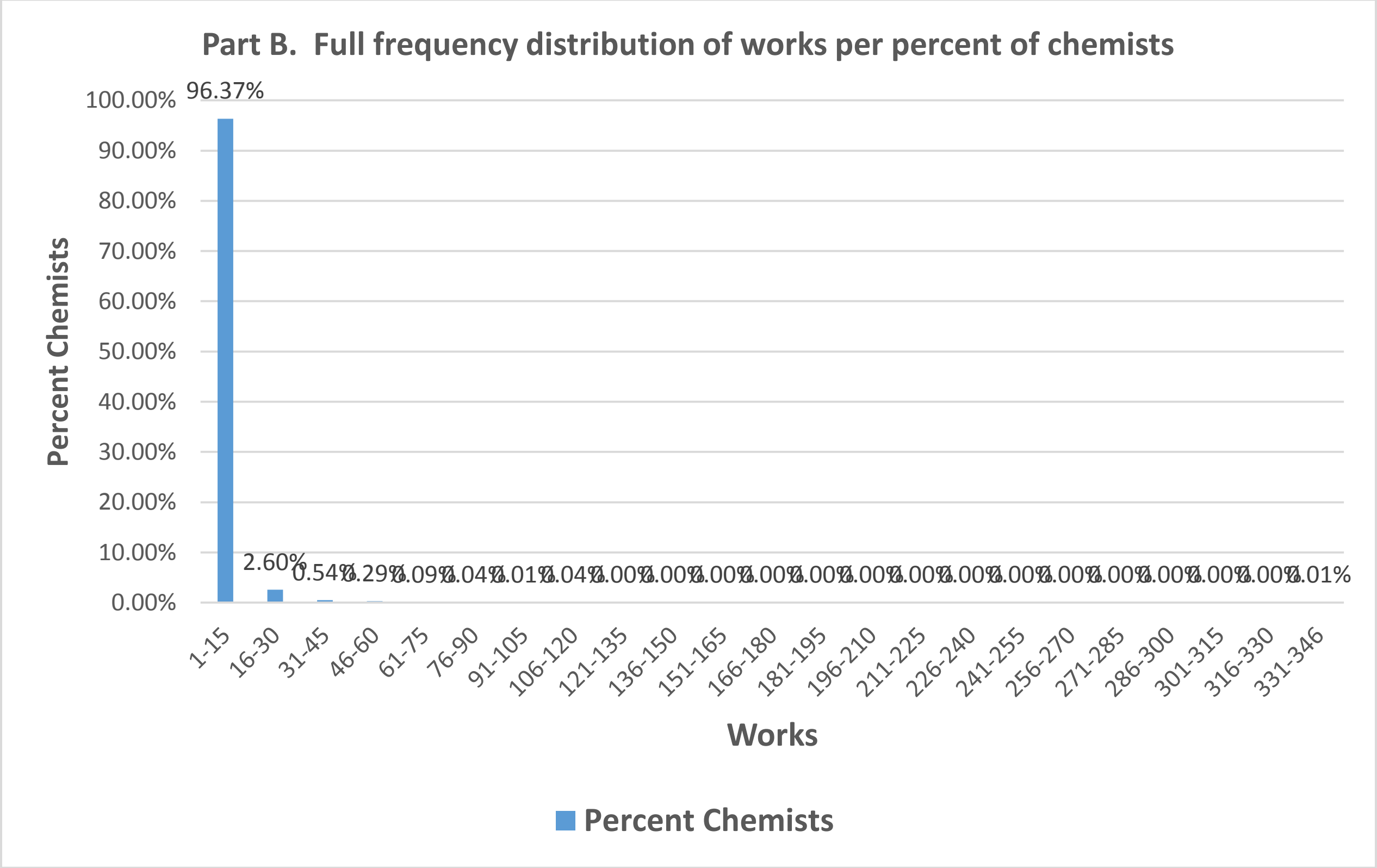

**Figure 3. Full frequency distributions of Lotka's Auerbach data: each bin approximating one-half of Lotka's right-truncation point of 17**

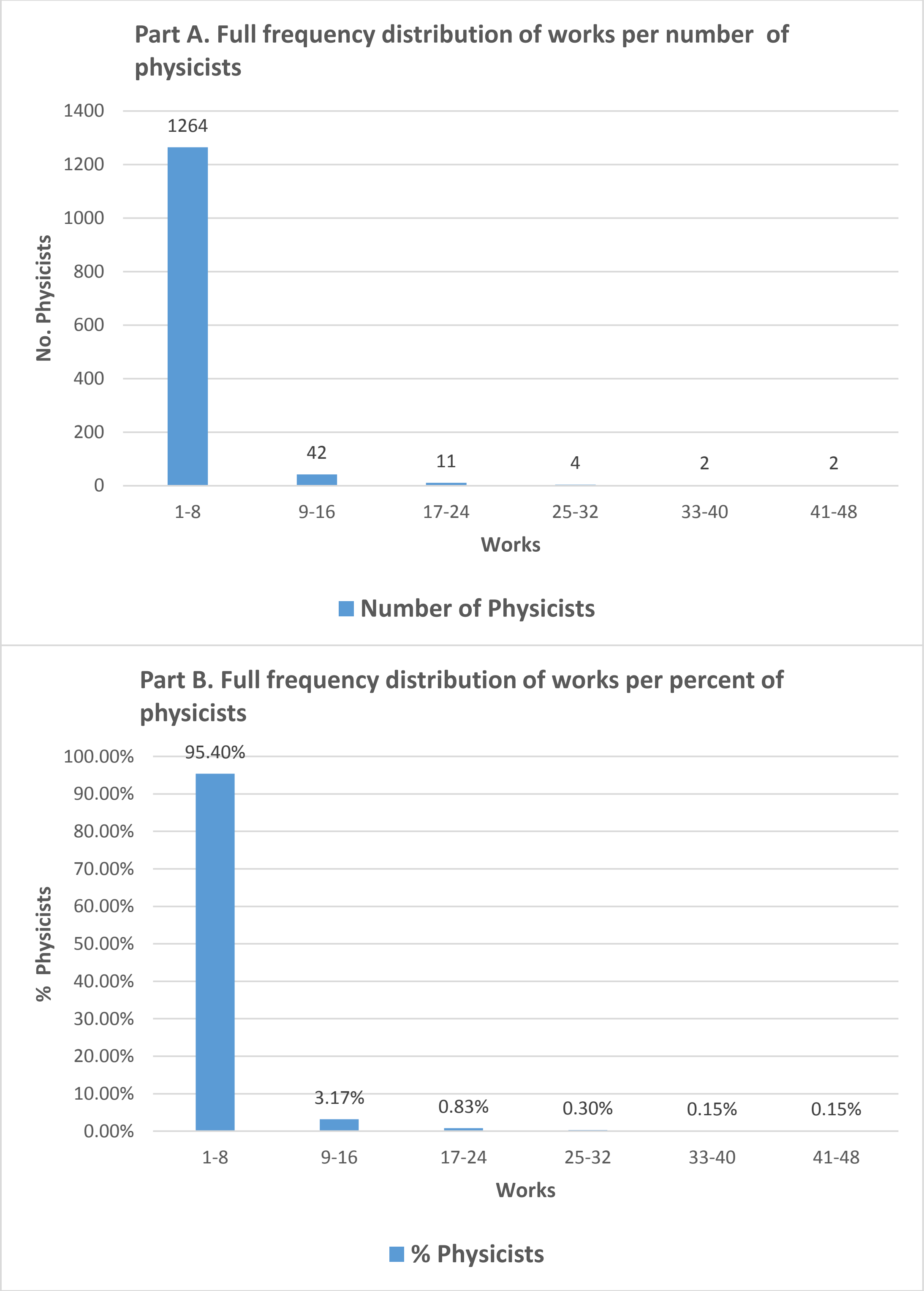

**Figure 4. Lotka's truncated distribution of his *Chemical Abstracts* data and his R^2 fit to the regression line on the log-log plot**

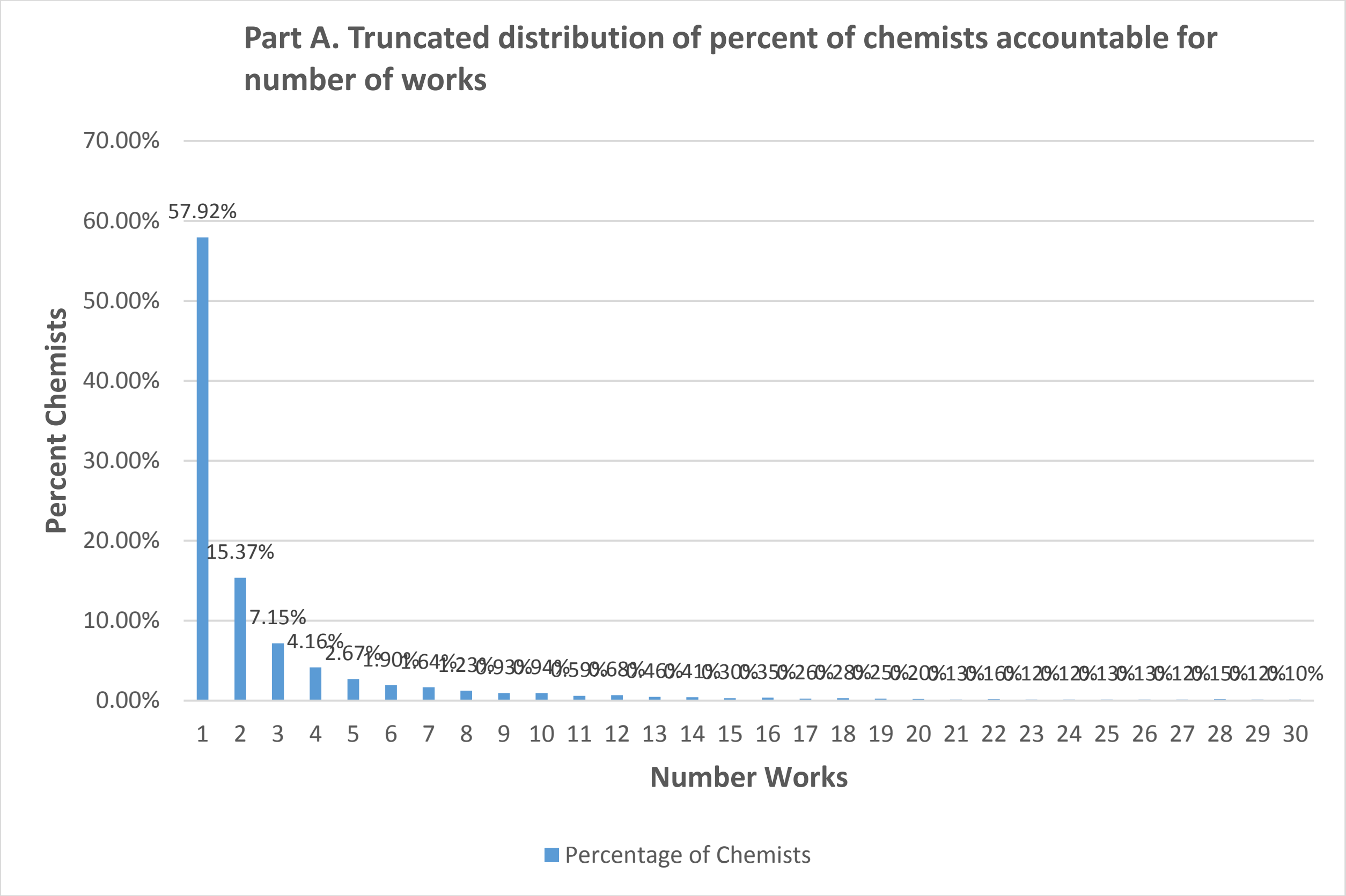

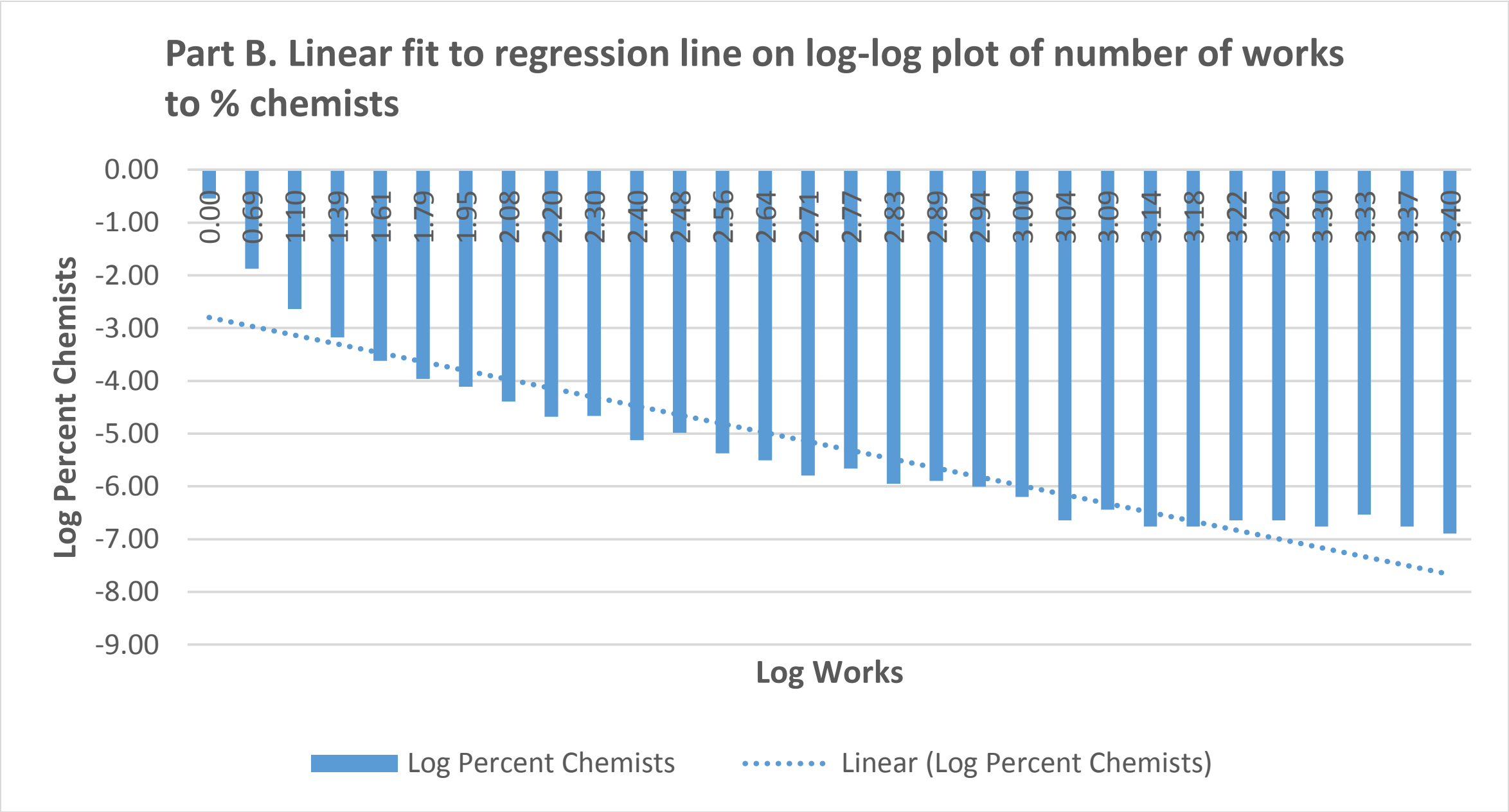

**Slope/Exponent = 1.888. R^2 = 0.99. F = 3676.9. Degrees of Freedom = 28**

**Figure 5. Lotka's truncated distribution of his Auerbach and his R^2 fit to the regression line on the log-log plot**

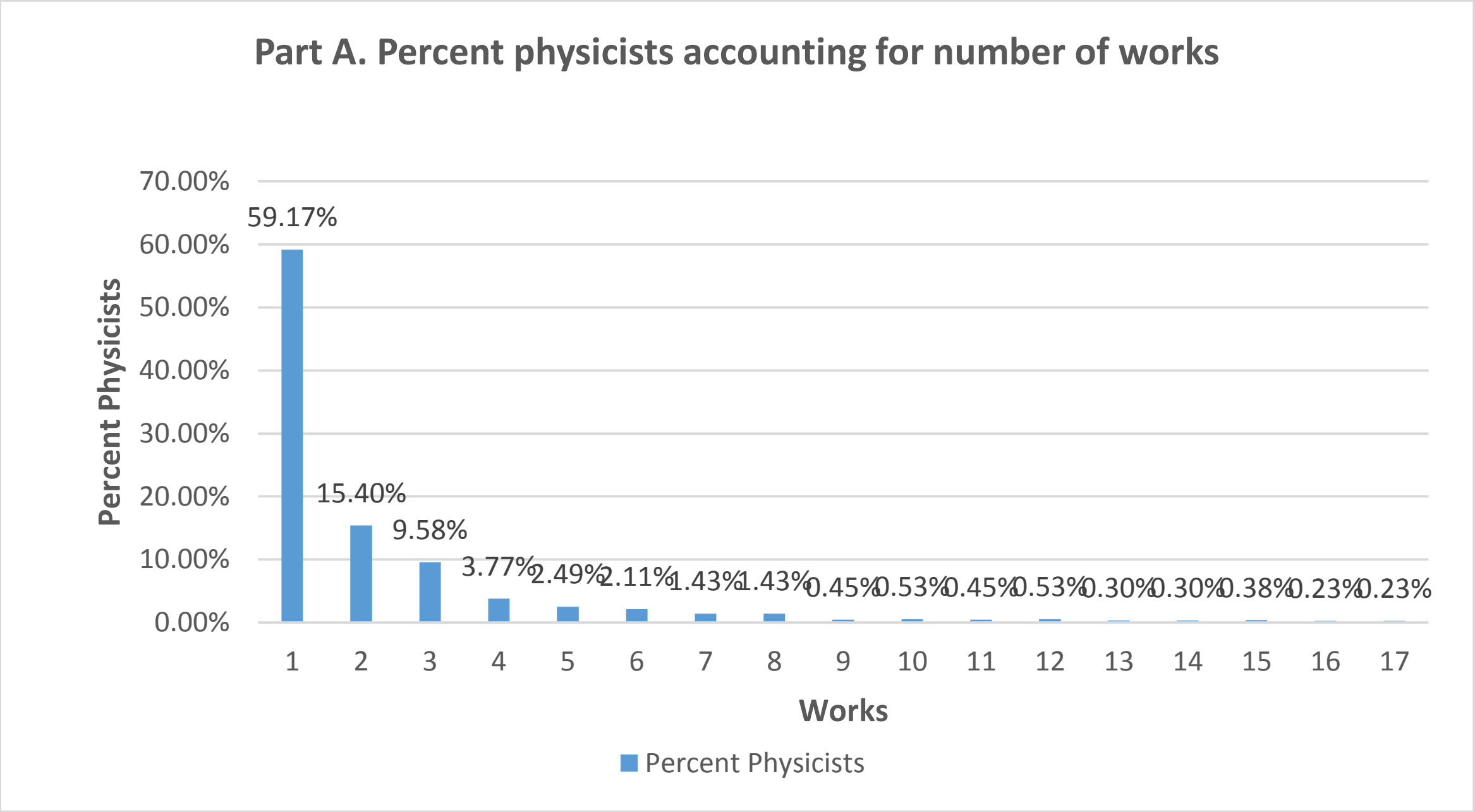

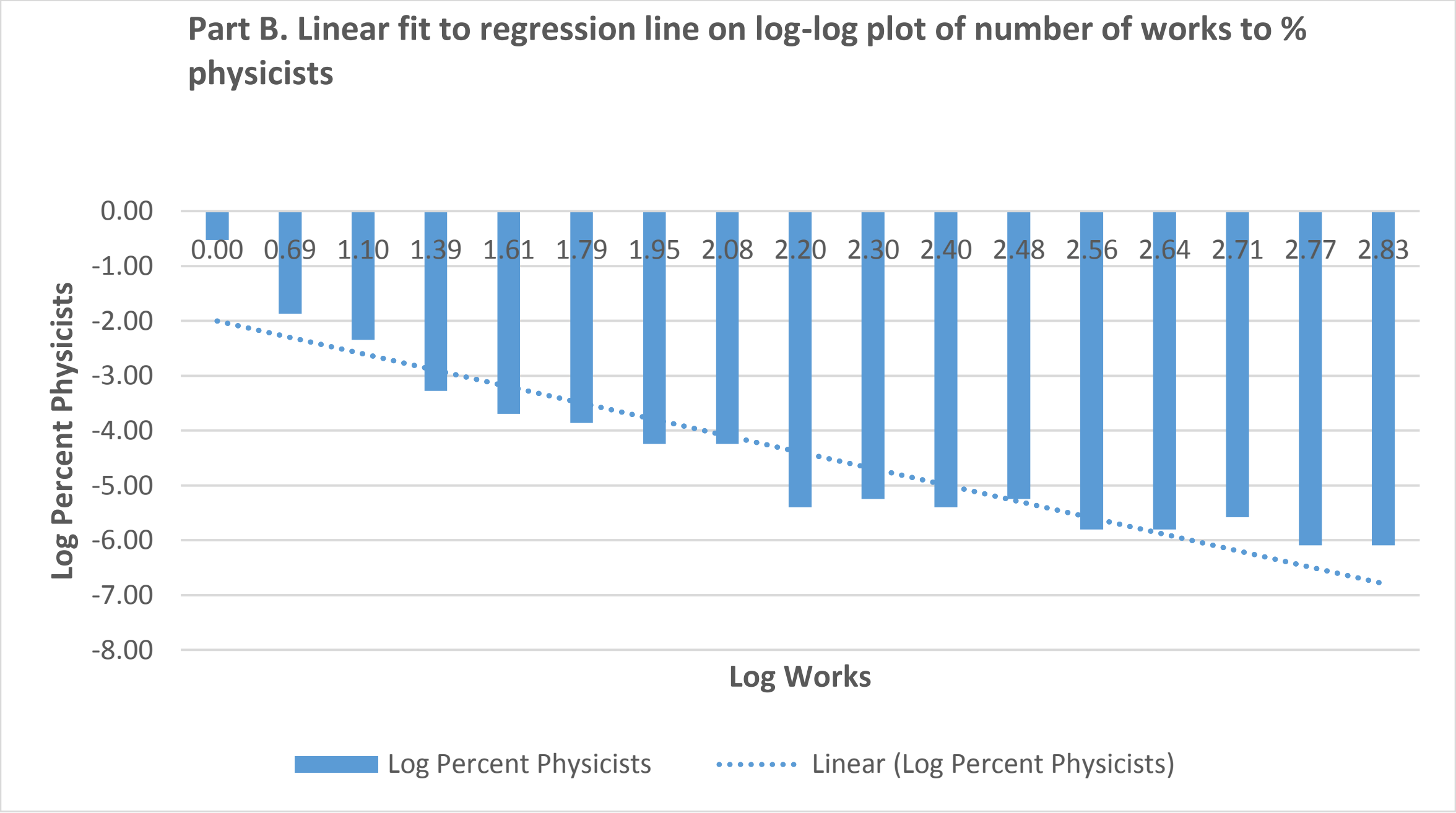

**Slope/Exponent = 2.021. R^2 = 0.98. F = 763.8. Degrees of Freedom = 15**

equivalent to maximizing R^2. Figures 4 and 5 above demonstrate with Excel Lotka's method of obtaining the R^2 fit to the trendline or regression line on his truncated distributions of his *Chemical Abstracts* and Auerbach data. The *Chemical Abstracts* distribution is exemplified in Figure 4, and the Auerbach distribution, in Figure 5. In each figure Part A shows the distribution, and Part B replicates the method of obtaining the R^2 fit to the trendline on the log-log plot with the Excel's LINEST function. To understand these figures, one has to understand two things. First, in this part of the paper Lotka measured number of authors only in percentages. Second, Lokta plots the number of papers (n) on the x-axis and the percentage of authors with n papers on the y-axis. Hence the y-axis shows the probability that a randomly selected author from the list will have exactly n papers. For example, in his *Chemical Abstracts* data 1 paper is considered as producing 3,991 chemists or 57.9%. Therefore, in both Parts B the log of percentage of scientists is the dependent variable, and the log of the number of papers is the independent variable. .

With this information the figures should be self-explanatory, and it is necessary only to emphasize the main result. In both cases the absolute value of the slope is equal to the exponent, and in both cases the slope/exponents are equal to the third decimal place the exponents Lotka utilized to calculate his law—1.888 for his *Chemical Abstracts* data and 2.021 for his Auerbach data. Given the passage of time and the advance in technology, such an outcome seems almost miraculous, and it proves one thing—Lotka based the derivation of his inverse square law of scientific productivity on the very method of identifying power-law behavior by the R^2 fit to the regression line on the log-log plot that Clauset, Shalizi, and Newman (2009) now consider unreliable.

**References**

Clauset, A., Shalizi, C. R., & Newman, M. E. J. (2009). Power-law distributions in empirical data. *SIAM Review*, 51 (4), 661-703.

Lotka, A. J. (1926). The frequency distribution of scientific productivity. *Journal of the Washington Academy of Sciences*, 16 (12), 317-322.

Newman, M. E. J. (2005). Power laws, Pareto distributions and Zipf's law. *Contemporary Physics*, 46 (5), 323-351.